\documentclass[11pt]{article}
\usepackage[margin=2.5cm]{geometry}
\usepackage{graphicx}
\usepackage{booktabs}
\usepackage{tabularx}
\usepackage{xcolor}
\usepackage{enumitem}
\usepackage{tikz}
\usetikzlibrary{arrows.meta,positioning,calc,shapes.geometric}
\usepackage{amsmath}
\usepackage{hyperref}
\hypersetup{colorlinks=true,linkcolor=blue,citecolor=blue,urlcolor=blue}
\begin{document}

\title{Modeling and Simulation Based Engineering \\ in the Context of Cyber-Physical Systems \\ \large Execution Semantics under Physical Constraints}
\author{Alexandre Muzy \\[6pt]
\small International Laboratory on Learning Systems (ILLS), CNRS, McGill University, \'ETS, \\
\small MILA, Universit\'e Paris-Saclay, Montr\'eal, Qu\'ebec, Canada. \\
\small \texttt{alexandre.muzy@cnrs.fr}}
\date{}
\maketitle

\begin{abstract}
Cyber-Physical Systems (CPS) produce behavior through execution on substrates coupling computation with physical processes. However, usual engineering approaches do not treat execution semantics as first-class engineering entities. Formal verification reasons about model behaviors under fixed semantic assumptions that are not revisable and do not account for physical execution constraints. Simulation-based validation explores scenarios under execution semantics that are implicitly determined by the simulation engine. In both cases, physical constraints of the execution substrate are addressed as implementation details rather than as semantic boundary conditions. In this article, it is hypothesized that making execution semantics explicit as first-class engineering entities is necessary and sufficient to bridge the gap between verified model behaviors and validated executed behaviors in CPS. To test this hypothesis, Modeling and Simulation Based Engineering (MSBE) is proposed: a methodology grounded in the Theory of Modeling and Simulation. MSBE formalizes execution conditions as four components: execution semantics, activity (behaviorally meaningful changes), admissibility constraints (physical bounds), and specified properties (behavioral guarantees). MSBE organizes engineering around an iterative cycle alternating formal execution, experimental execution, verification, and activity-mediated validation. Executability is defined as stabilization of execution conditions and the induced admissible model space. The cycle is applied to four CPS classes (human-centric, biophysical, technological, and digital twins). These applications show that the framework generalizes beyond CPS to any system whose behavior depends on explicitly defined execution conditions.
\paragraph{Keywords:}Modeling and Simulation Based Engineering \and Cyber-Physical Systems \and Executability \and Execution Semantics \and Physical Constraints \and Activity
\end{abstract}

\section{Introduction}
\label{sec:intro}

Across engineering domains, Cyber-Physical Systems (CPS) are increasingly embedded, ranging from automated vehicles to large-scale sensing and actuation infrastructures for health, agriculture, and environmental monitoring \cite{lee2008cps,rajkumar2010cps}. In CPS, software-driven computations are tightly coupled with physical processes through sensing and actuation. System-level behavior emerges from execution over time across multiple substrates: the computational platforms executing software and the physical media in which dynamics unfold. For instance, in a smart grid, system behavior is produced by the real-time execution of distributed control software interacting with physical power flows.

Despite the increasing prevalence of CPS, the rigorous Verification and Validation (V\&V) of their behavior remains challenging. In this paper, \textit{behavioral verification} refers to checking whether a model admits behaviors satisfying specified properties, including safety, liveness, well-definedness, temporal constraints, causal ordering, stability, and consistency across executions \cite{alur2015cps,platzer2018logical}. \textit{Behavioral validation} refers to comparing the behavior observed during system execution with the expected behavior \cite{balci1998vv,zeigler2018tms}.

Verification approaches make it possible to prove behavioral properties on simplified or abstract models. Simulation-based validation approaches allow the comparison of experimental and simulated behaviors, albeit without formal guarantees on behavioral coverage or equivalence. However, both face fundamental limitations in the context of CPS. Existing formal verification approaches \cite{alur2015cps,henzinger1996hybrid,platzer2018logical} effectively address behavioral properties at the model level, but are primarily applicable to closed CPS, where behavior is assumed to be fully determined by the model and its formal execution semantics. As a result, they verify properties under semantic assumptions that are fixed at specification time, without formalizing the conditions under which these guarantees transfer to execution on physically constrained substrates. Simulation- and co-simulation-based validation approaches \cite{gomes2018cosimulation,cremona2019hybrid} support the exploration and testing of execution scenarios, but remain limited to scenario-based coverage. They do not provide a specification of emergent behavior resulting from interactions between CPS components. The effective execution order is often implicitly determined by the simulation engine or the underlying execution platform.

To overcome these V\&V limitations, we propose an explicit engineering of execution semantics to relate verified model behaviors to validated executed behaviors in CPS. We call this approach Modeling and Simulation Based Engineering (MSBE).

MSBE extends the Theory of Modeling and Simulation (TMS) \cite{zeigler1976tms,zeigler2018tms} to engineering and execution. In TMS, a model is not defined solely by its static mathematical structure but also by its associated simulation, understood as the ordered computation of state transitions over time. A simulation produces a trace representing how the model states evolve according to its assumed dynamics. The behavioral meaning of a model is therefore inseparable from the way its state transitions are computed. MSBE generalizes this perspective beyond simulation through the notion of \emph{execution}: the realization of model state transitions on a given substrate. Such a substrate may be digital (computers, embedded processors), physical (a robotic system), or hybrid, as in CPS where computational substrates are coupled in closed loop with physical substrates.

\emph{Execution semantics} specify explicitly how model states evolve over real time during execution: admissible state transitions, update orderings, timing assumptions, and the causal relations between internal computations and external interactions. Execution semantics determines how behavior is effectively produced, independently of a specific simulation engine or platform. Execution on a given substrate is inherently subject to \emph{physical constraints} arising from the execution substrate and its interaction with the physical world. These constraints include latency, update rates, bandwidth, energy availability, actuation precision, etc. They bound the temporal and causal validity of execution and restrict the set of admissible behaviors. In CPS, simulation can then be understood as a particular form of execution in which state transitions are realized under controlled semantic assumptions.

In this context, \emph{activity} provides a behavioral measure capturing meaningful changes produced by execution \cite{muzy2019exploiting}. Rather than tracking all state transitions, activity quantifies those that produce observable behavioral differences. It provides the semantic granularity at which behavioral admissibility is evaluated and serves as a link between discrete computations, continuous dynamics, and physical constraints. Formal definitions of activity measures for continuous, hybrid, and discrete-event traces are provided in Appendix~\ref{app:activity}.

\emph{Executability} denotes the ability of a model-based system to sustain coherent execution over time under physical constraints. It goes beyond mere runnability. Executability characterizes the capacity to preserve admissible and meaningful activity as execution conditions evolve. This notion is central to CPS engineering, where long-running, interactive, and physically grounded executions are the norm.

MSBE contributes to CPS engineering along several dimensions. Conceptually, it shifts the engineering focus from models to behaviors produced via execution under physical constraints, and provides explicit execution semantics independent from specific tools. Methodologically, it bridges verified model behaviors, simulated behaviors, and observed execution behaviors, reducing the semantic gap in V\&V. It integrates physical constraints at the semantic level and uses activity as a unifying behavioral measure. From an engineering perspective, it captures sustained behavioral coherence through the notion of executability and applies across CPS domains including autonomous systems, smart infrastructures, robotics, and digital twins. The framework also generalizes beyond CPS, as discussed in Section~\ref{subsec:cps-synthesis}.

Section~\ref{sec:limits} reviews the limits of current software and systems engineering practices for CPS.
Section~\ref{sec:cps-msbe} reframes CPS as execution-constrained systems and introduces the MSBE iterative executability cycle.
Section~\ref{sec:classes} discusses representative CPS classes and the distinct execution challenges they raise.
Section~\ref{sec:roadmap} derives implications and sets a research roadmap covering executability formalization, constraint-aware semantics, verification-validation bridging, and integration with language-level frameworks.
Section~\ref{sec:conclusion} concludes and draws perspectives.

\section{Limits of Model-based Engineering Approaches for Cyber-Physical Systems: a Historical Survey}
\label{sec:limits}

This historical survey emphasizes how behavioral semantics progressively became more explicit across model-based engineering approaches, while remaining partially implicit with respect to execution under physical constraints.

In 1967, A.W. Wymore provided a mathematical theory of Systems Engineering \cite{wymore1967}. Systems were defined as inputs–states–outputs with formal specifications. This system-theoretic framework introduced structured decomposition and the separation between specification, design, and implementation steps.

In 1976, B.P. Zeigler complemented Wymore's framework by formalizing simulation semantics in the Theory of Modeling and Simulation (TMS) \cite{zeigler1976tms}. He proposed unified mathematical structures for computational models of systems, making models simulable, analyzable, and composable across domains. A simulation-oriented formalism, Discrete Event System Specification (DEVS), provided a rigorous semantic framework for specifying and simulating discrete-event system behavior.

In 1993, A.W. Wymore introduced Model-Based Systems Engineering (MBSE) \cite{wymore1993} in industrial systems engineering to organize requirements, architecture, analysis, and verification around shared system models across the lifecycle.

In 1997, in Software Engineering, executable object modeling with statecharts was proposed as a means to provide operational semantics to behavioral models \cite{harel1997statecharts}.

In the late 1990s, Model-Based Design (MBD) \cite{kokar1999mbd} was explicitly discussed as a methodology applied to the design of embedded software, particularly in control-oriented and real-time systems, relying on executable models and tool-supported simulation.

In 2004 and 2005, Model-Driven Engineering (MDE) \cite{bezivin2005unification,mellor2004mda} further generalized the role of models by promoting them as primary engineering artifacts for specification, transformation, and code generation across software development processes.

In 2006, Discrete Event System Specification (DEVS) and its Dynamic Structure extension (DSDEVS) \cite{barros1997modeling} were employed to provide formal simulation semantics to metamodels describing system behavior. In particular, Muzy et al. \cite{muzy2006} proposed a framework for the visual specification and simulation of cellular systems in which the simulation semantics of the behavioral metamodel were defined in DEVS. In 2007, in a language for modeling agents embedded in a physical environment \cite{muzy2007}, the simulation semantics relied on Dynamic Structure DEVS (DSDEVS), enabling structural changes during simulation. In both cases, DEVS-based formalisms were used to endow visual or metamodel-based specifications with precise operational simulation semantics.

More recently, Modeling and Simulation-Based Systems Engineering (MSBSE) aims at combining MBSE and DEVS \cite{gianni2018modeling,zhang2023modeling}. DEVS provides a formal framework for specifying discrete-event system behavior together with its associated simulation semantics. However, in its core discrete-event formulation, DEVS does not directly provide abstract mathematical reasoning principles for hybrid behaviors combining continuous and discrete dynamics typical of CPS without additional hybridization assumptions.

Formal approaches to hybrid systems, e.g., hybrid automata and related model-checking techniques \cite{henzinger1996hybrid,alur2015cps,platzer2018logical}, address specific verification problems under strong modeling assumptions. While these approaches provide rigorous behavioral analysis frameworks, they do not constitute a unifying engineering methodology for executable system semantics under physical constraints.

This motivated the development of a more abstract hybrid formalism within the Iterative Specification (IterSpec) framework \cite{muzy2017iterative}, later consolidated in the advanced version of TMS \cite{zeigler2018tms}. IterSpec is a formal mechanism by which admissible system behaviors are progressively refined through successive behavioral constraints until semantic stabilization is achieved. A key result in TMS \cite{zeigler2018tms} (Ch.~10) shows that Finite and Timed Parallel DEVS can simulate any iterative specification, providing a computational realization of the formalism and a bridge toward DEVS-based model checking. Within this formal setting, it becomes possible to reason about hybrid cyber–physical behaviors at the specification level, providing a foundation for their subsequent engineering realization.

In conclusion, traditional model-based engineering approaches primarily focus on structural specification and lifecycle organization, while executable modeling approaches such as statecharts and Model-Based Design provide operational behavioral semantics at the software or tool level. However, TMS and DEVS extend this perspective by defining behavior through explicit and formal simulation semantics. For cyber-physical systems, this distinction becomes critical, as structural modeling or software-level executability alone is insufficient to reason rigorously about behavior emerging from execution under physical constraints. This historical evolution therefore reveals the need for an engineering framework in which execution semantics under physical constraints are treated as first-class design artifacts.

\subsection{Positioning with respect to related work}
\label{subsec:positioning}

Beyond the historical evolution traced above, several contemporary research directions address aspects of executable modeling, co-simulation, and CPS verification. This subsection positions MSBE with respect to these contributions along three themes.

\paragraph{Executability of models.}

The execution of modeling languages has been extensively studied in the MDE community. Ciccozzi et al.\ \cite{ciccozzi2019execution} provide a systematic review of 63 research studies and 19 tools for executing Unified Modeling Language (UML) models, distinguishing translational and interpretive strategies. At the standards level, Foundational UML (fUML) defines a precise operational semantics for a subset of UML, and xMOF \cite{mayerhofer2013xmof} extends this approach to arbitrary Domain-Specific Modeling Languages (DSMLs) by providing executable semantics based on the Meta-Object Facility. These standards make execution semantics explicit at the language level: they specify how language constructs are interpreted. The GEMOC initiative (Globalized use of Models of computation for the Engineering of Complex systems) \cite{combemale2014globalizing} goes further by addressing the composition of heterogeneous execution semantics for DSMLs. GEMOC provides a framework in which multiple behavioral semantics can coexist and interact within a single model, enabling the coordination of discrete, timed, and concurrent execution regimes at the language level. The Ptolemy framework \cite{eker2003ptolemy} addresses a related concern at the simulation level by defining models of computation that govern how components interact under different execution regimes (dataflow, synchronous, discrete-event, continuous-time). An industrial survey by Liebel et al.\ \cite{liebel2018sosym} documents the state of practice in model-based engineering for embedded systems and identifies persistent gaps between modeling and deployment. The challenges of verifying and validating SysML models in industrial contexts are further documented by Baduel et al.\ \cite{baduel2018sysml}, who report on the difficulty of ensuring model correctness without domain-specific execution semantics.

These approaches define executability at the level of the modeling language or the simulation tool. fUML and xMOF specify how a model is interpreted. GEMOC specifies how heterogeneous language semantics are composed. Ptolemy specifies how heterogeneous simulation components interact. The question they address is: can this model be executed, and under which language-level or tool-level semantics? MSBE operates at a different level. It asks: given that the model is executable under specified language semantics, does the behavior produced by execution remain admissible when realized on a physically constrained substrate? The execution strategies classified by Ciccozzi, the language-level semantics defined by fUML/xMOF, the composition framework of GEMOC, the models of computation of Ptolemy, and the tool-level practices documented by Liebel all provide execution mechanisms. MSBE provides the framework to evaluate whether these mechanisms produce admissible behaviors under physical constraints and to iteratively refine execution conditions when they do not. The two levels are complementary: language-level executability is a prerequisite for substrate-level executability.

\paragraph{Co-simulation and heterogeneous execution semantics.}

Co-simulation has become a standard approach for exploring CPS behavior by composing heterogeneous simulators. Gomes et al.\ \cite{gomes2018cosimulation} provide a comprehensive survey covering master algorithms, coupling strategies, and the Functional Mock-up Interface (FMI) standard. Cremona et al.\ \cite{cremona2019hybrid} address the specific problem of time management in hybrid co-simulation, showing that temporal consistency requires explicit treatment. Recent work published in SoSyM has explored higher-level semantic annotations for Functional Mock-up Unit (FMU) composition \cite{rindaroy2025fmu}.

In co-simulation, the effective execution order is determined by the master algorithm, step sizes, and interpolation strategies. These choices constitute implicit execution semantics. MSBE makes this observation explicit: co-simulation is a form of formal execution whose execution semantics are instantiated by the master algorithm. The contribution of MSBE is to compare these specified semantics with the semantics induced by experimental execution on a target substrate, and to iteratively refine execution conditions when they diverge.

\paragraph{Verification and analysis of CPS.}

The foundational challenges of CPS engineering have been articulated by Lee \cite{lee2008cps} and Rajkumar et al.\ \cite{rajkumar2010cps}, who argue that current computing abstractions are inadequate for systems tightly coupling computation and physical processes. Alur \cite{alur2015cps} provides a rigorous textbook treatment covering model checking, temporal logic, hybrid systems, and real-time scheduling. Platzer \cite{platzer2018logical} develops differential dynamic logic for deductive verification of hybrid programs, with tool support through KeYmaera~X. Hybrid automata and associated model-checking techniques \cite{henzinger1996hybrid} address specific verification problems under strong modeling assumptions. Model checking of behavioral specifications, including techniques for managing partial models and inconsistency \cite{chechik2003modelchecking}, provides the formal machinery for reasoning about system properties at the model level. In the synchronous paradigm, Berry's foundational work on Esterel \cite{berry2000esterel} demonstrates that making execution semantics deterministic at the language level enables rigorous verification. This has been applied to biophysical CPS: De Maria et al.\ \cite{demaria2016lustre} use the synchronous language Lustre to model neuronal archetypes as reactive systems and verify their temporal properties through model checking. At the level of distributed software systems, Güdemann et al. \cite{poizat2016verchor} address the conformance of peer implementations to choreography specifications, verifying whether distributed behaviors match a global behavioral contract.

These approaches reason about the possible behaviors of formal models under their defined semantics. They provide the formal machinery for the verification branch of the MSBE cycle. The synchronous approach of Berry and De Maria et al.\ makes execution semantics explicit at the language level; Poizat et al.\ verify behavioral conformance of distributed compositions; Chechik et al.\ provide model-checking foundations for behavioral properties. However, these approaches do not focus on the relation between verified model behaviors and behaviors produced by execution on physical substrates. MSBE complements them by adding the experimental branch, formalizing validation as a comparison between specified and experimentally induced execution conditions, and using activity as the mediator between the two.

Recent work in SoSyM journal on digital twins \cite{gil2025dt,mertens2025dartwin,mertens2025dtvalidation} proposes architectures for coupled digital-physical systems, semantic lifting, and continuous validation. These contributions address architectural and notational concerns. MSBE provides the formal underpinning: Section~\ref{subsec:digital-twin} instantiates the MSBE cycle on digital twins, formalizing the conditions under which synchronization remains admissible and executability is sustained over operational lifetime.

\paragraph{Summary.}
 
No existing approach simultaneously addresses the three gaps identified in this survey: (i) making execution semantics explicit as engineering entities independent of specific tools, (ii) integrating physical constraints at the semantic level, and (iii) formalizing the articulation between model-level verification and execution-level validation. MSBE addresses these three gaps through explicit execution conditions, the separation between formal and experimental execution, activity as mediator, and executability as stabilization. Section~\ref{sec:cps-msbe} formalizes these concepts.

\section{Cyber-Physical Systems as Execution-Constrained Systems}
\label{sec:cps-msbe}

Cyber-physical systems (CPS) are not merely software-intensive systems. They are execution-constrained systems whose behavior emerges from the interaction between computational dynamics and physical processes.
Their engineering therefore requires explicit reasoning about execution under physical constraints.

The key concepts underlying MSBE were introduced in Section~\ref{sec:intro}. This section reframes them for CPS and develops the formal cycle. Physical constraints in CPS are not design choices but boundary conditions imposed by the execution substrate and its coupled physical environment. They are not secondary to performance: they restrict the set of admissible behaviors and must therefore be represented at the semantic level. Simulation can be understood as a controlled form of execution under idealized assumptions; real execution introduces additional constraints that may alter behavioral admissibility.

Activity plays a dual role in this context. Formally, it provides a measure that reduces the admissible behavior space by quantifying which state transitions are behaviorally relevant. Operationally, it corresponds to observable behavior produced during execution under physical constraints. By focusing on activity rather than raw state updates, engineering reasoning can relate modeling assumptions, architectural structure, and validation evidence to execution traces in a domain-independent manner.

Within advanced TMS, the mechanism of \emph{iterative system specification} \cite{muzy2017iterative,zeigler2018tms} provides a formal means to progressively refine the admissible behavior space of a system through successive behavioral constraints. In the context of CPS, admissibility depends on execution under physical constraints. Observed activity during execution may reveal behaviors that violate intended constraints or expose implicit assumptions. This leads to refinement of behavioral constraints, adjustments of the model's mathematical structure, or modified execution semantics. The classical analysis--design--validation cycle can therefore be interpreted as the operational manifestation of this underlying semantic refinement mechanism.

\subsection{Execution-centered iterative cycle}

\begin{figure}[t]
\centering
\begin{tikzpicture}[
    node distance=0.5cm,
    every node/.style={align=center},
    formal/.style={draw=blue!70, thick, rounded corners, fill=blue!10, minimum width=4.6cm, minimum height=1cm},
    experimental/.style={draw=orange!80!black, thick, rounded corners, fill=orange!15, minimum width=4.6cm, minimum height=1cm},
    neutral/.style={draw=black, thick, rounded corners, fill=gray!10, minimum width=5cm, minimum height=1cm},
    decision/.style={draw=black, thick, diamond, aspect=2.1, inner sep=2pt},
    arrow/.style={->, thick}
]

\node[formal] (model)
{\textbf{Specify execution conditions} $EC_i=(\Sigma_i,\alpha_i,\varphi_i,P_i)$\\
\textbf{Specify validation criteria} $C_i$\\
\textbf{Specify model} $M_i \in \mathcal{M}_i$ under $EC_i$};

\node[formal, below left=0.3cm and -2.2cm of model] (fexec)
{\textbf{Formal execution}\\
$b_i=\mathrm{Exec}_f(M_i,\Sigma_i)$};

\node[experimental, below right=0.3cm and -2.2cm of model] (eexec)
{\textbf{Experimental execution}\\
$\widehat{b}_i=\mathrm{Exec}_e(M_i)$\\
$\widehat{EC}_i=(\widehat{\Sigma}_i,\widehat{\alpha}_i,\widehat{\varphi}_i)$};

\node[formal, below=of fexec] (verif)
{\textbf{Verification}\\
$M_i,EC_i \models P_i$};

\node[experimental, below=of eexec] (valid)
{\textbf{Validation}\\
$EC_i$ vs $\widehat{EC}_i$\\
according to $C_i$};

\node[neutral, below=3.3cm of model] (assess)
{\textbf{Execution Assessment}\\
Sustained execution?};

\node[decision, below=of assess] (dec)
{Stabilized?};

\node[formal, below left=0.8cm and 0cm of dec] (refine)
{\textbf{Iterative Specification}\\
$EC_{i+1}=\mathrm{Specify}(EC_i,\widehat{EC}_i,C_i)$\\
$\mathcal{M}_{i+1}\subseteq\mathcal{M}_i,\;\; M_{i+1}\ \text{under}\ EC_{i+1}$};

\node[below right=0.5cm and 0cm of dec] (execb)
{\textbf{Executability}\\
(fixed point)\\
$EC_{i+1}=EC_i,\;\mathcal{M}_{i+1}=\mathcal{M}_i$};

\draw[arrow] (model) -- (fexec);
\draw[arrow] (model) -- (eexec);

\draw[arrow] (fexec) -- (verif);
\draw[arrow] (eexec) -- (valid);

\draw[->, thick, rounded corners=6pt] (verif) |- (assess);
\draw[->, thick, rounded corners=6pt] (valid) |- (assess);

\draw[arrow] (assess) -- (dec);

\draw[arrow] (dec) -- node[left]{no} (refine);
\draw[arrow] (dec) -- node[right]{yes} (execb);

\draw[->, thick, rounded corners=6pt] (refine) -- ++(-3.4,0) |- (model);

\node[anchor=south east] at ($(current bounding box.south east)+(0.4cm,1.4cm)$) {
\begin{tikzpicture}[scale=0.8]
\node[formal, minimum width=0.4cm, minimum height=0.4cm] (lf) {};
\node[right=0.1cm of lf] {\scriptsize Formal};

\node[experimental, minimum width=0.4cm, minimum height=0.4cm, below=0.2cm of lf] (le) {};
\node[right=0.1cm of le] {\scriptsize Experimental};
\end{tikzpicture}
};
\end{tikzpicture}
\caption{Execution-centered MSBE cycle. Blue: formal domain. Orange: experimental domain. $EC_i = (\Sigma_i, \alpha_i, \varphi_i, P_i)$ is specified; $\widehat{EC}_i = (\widehat{\Sigma}_i, \widehat{\alpha}_i, \widehat{\varphi}_i)$ is derived from observed traces. Properties $P_i$ transfer when validation confirms that execution conditions are preserved. $C_i$: validation criteria. Executability: stabilization of $EC_i$ and $\mathcal{M}_i$.}
\label{fig:msbe-cycle}
\end{figure}
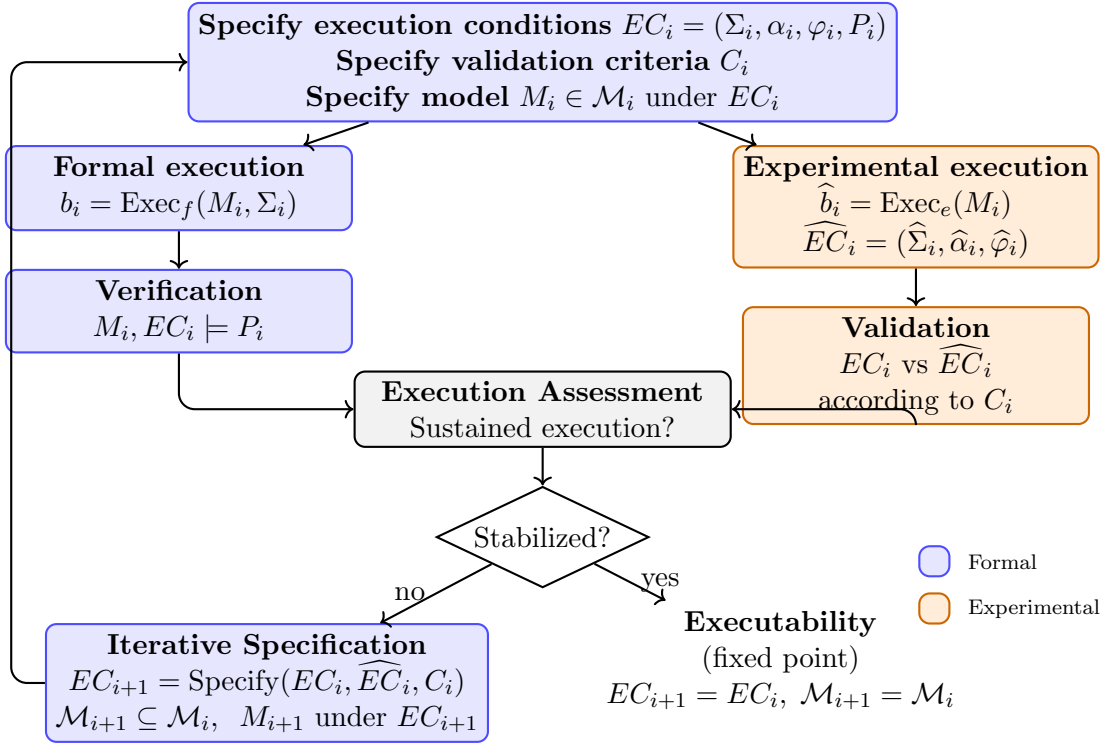

Figure~\ref{fig:msbe-cycle} illustrates the execution-centered
iterative cycle underlying MSBE.

MSBE applies whenever system behavior depends on explicitly defined
execution conditions. Rather than treating execution as an
implementation detail, the methodology makes execution semantics and
their associated constraints part of the engineering process itself.
This contrasts with classical development processes (e.g., the V-model  \cite{forsberg1991vmodel}),
which organize a sequential refinement of artifacts (models,
specifications, code) under implicitly assumed and stable execution
conditions. In such processes, verification checks consistency between
artifacts, and validation follows implementation under fixed runtime
assumptions. In contrast, MSBE recognizes that execution conditions
shape behavior and may evolve during engineering. They must therefore
be explicitly specified, evaluated, and, when necessary, revised. The
objective is not only to ensure artifact correctness, but to ensure
that system behavior remains coherent under its defined execution
conditions.

\noindent The cycle consists of the following steps.

\paragraph{Specification of Execution Conditions.}

The process begins by specifying formal Execution Conditions (EC)
derived from requirements. The subscript $i$ denotes the current
iteration of the cycle; at each iteration, execution conditions,
models, and the admissible model space may be revised:

\[
EC_i = (\Sigma_i, \alpha_i, \varphi_i, P_i),
\]

\noindent where:

\begin{itemize}

\item $\Sigma_i$ denotes \textit{execution semantics} defined at different
specification levels:
behavior level (ordering and triggering of input–output computations),
system level (state-space structure, transition rules, and temporal
progression), or
network-of-systems level (causality, communication, and
synchronization of interacting systems);

\item $\alpha_i$ denotes an \textit{activity measure} defined relative
to $\Sigma_i$,
\[
\alpha_i : \mathcal{B}_{\Sigma_i} \rightarrow \mathcal{A}_i,
\]
where $\mathcal{B}_{\Sigma_i}$ is the space of behavior traces realizable
under $\Sigma_i$, and $\alpha_i(b)$ captures computational behavior
segments (behavior level), state-transition events (system level), or
inter-system coordination events (network level).
The activity $\alpha_i$ is a measure that can be used as an abstraction.
It quantifies behaviorally relevant changes to 
match relevant state transitions.
The mapping from the trace space $\mathcal{B}_{\Sigma_i}$
to the activity space $\mathcal{A}_i$ reduces the dimensionality
of behavioral analysis while preserving the information needed
for constraint evaluation and validation.
The activity measure can take different forms depending on the nature
of the traces: the rate of continuous change (e.g., membrane potential
variation in a neuron), the total variation of a hybrid trace combining
continuous segments and discrete discontinuities, or the count or
frequency of discrete events (e.g., observed events, actuation
commands). These instantiations are formally defined in
Appendix~\ref{app:activity};

\item $\varphi_i$ denotes \textit{admissibility constraint functions}
on activity,
\[
\varphi_i : \mathcal{A}_i \rightarrow \mathcal{C}_i,
\]
where $\mathcal{C}_i$ represents constraint evaluations.
Following the iterative specification formalism \cite{muzy2017iterative,zeigler2018tms},
$\varphi_i$ operates at two levels:
on individual activity segments (e.g., timing bounds on a single
coordination delay, resource usage within one execution cycle),
and on their concatenation into behavior traces
(e.g., cumulative drift bounds, preservation of causal ordering
across successive segments, absence of temporal blocking).
A behavior trace is well-defined when all segments satisfy local
constraints and their concatenation satisfies global admissibility;

\item $P_i$ denotes specified structural, temporal, causal, and behavioral
\textit{properties} derived from requirements (e.g., safety, liveness,
stability), interpreted relative to the chosen execution semantics
level. Properties are verified on the model under specified execution
conditions. They are global by nature: they hold on behavior traces,
i.e., on the concatenation of activity segments. If the constraints
$\varphi_i$ are satisfied on segments and on their concatenation, and
the execution semantics $\Sigma_i$ are preserved, then verified
properties $P_i$ transfer to execution by construction.

\end{itemize}

Execution conditions define the semantic and constraint framework
within which models are specified and behaviors are realized.

Rather than comparing raw execution traces, MSBE aligns formal and
experimental executions through the activity measure.
Execution semantics generate behavior traces, while activity provides
a structured level at which constraint evaluation and validation are
performed. Activity therefore serves as the interface through which
behavioral coherence is assessed.

\vspace{1ex}
\paragraph{Specification of the Model.}

Given execution conditions $EC_i$, a model is specified:
\[
M_i \in \mathcal{M}_i
\quad \text{such that} \quad
M_i \text{ is intended to satisfy } EC_i.
\]

The model is thus defined under explicit execution conditions.

\vspace{1ex}
\paragraph{Formal and Experimental Execution.}

Formal execution (simulation) produces:

\[
b_i = \mathrm{Exec}_f(M_i, \Sigma_i),
\qquad
b_i \in \mathcal{B}_{\Sigma_i}.
\]

Experimental execution produces:

\[
\widehat{b}_i = \mathrm{Exec}_e(M_i).
\]

From $\widehat{b}_i$, one derives:

\[
\widehat{\alpha}_i = \alpha_i(\widehat{b}_i),
\qquad
\widehat{\varphi}_i = \varphi_i(\widehat{\alpha}_i),
\qquad
\widehat{\Sigma}_i = \mathrm{Infer}(\widehat{b}_i),
\]

where:
\begin{itemize}
\item $\widehat{\alpha}_i = \alpha_i(\widehat{b}_i)$:
the specified activity measure applied to the experimental trace,
producing segments extracted from observed data;
\item $\widehat{\varphi}_i = \varphi_i(\widehat{\alpha}_i)$:
the specified constraint functions applied to the experimentally
derived activity, evaluated on individual segments and on their
observed concatenation;
\item $\widehat{\Sigma}_i = \mathrm{Infer}(\widehat{b}_i)$:
the execution semantics inferred from the temporal and causal
structure of the experimental trace. Unlike $\widehat{\alpha}_i$
and $\widehat{\varphi}_i$, which are computed by applying specified
functions to experimental data, $\widehat{\Sigma}_i$ is reconstructed
from observed orderings, timings, and causal dependencies.
\end{itemize}

\noindent Experimental execution conditions are thus defined as

\[
\widehat{EC}_i =
(\widehat{\Sigma}_i,
\widehat{\alpha}_i,
\widehat{\varphi}_i).
\]

Properties $P_i$ are not evaluated directly on experimental traces.
Instead, validation assesses whether the execution conditions under
which $P_i$ was formally verified are preserved in experiment.
If $\Sigma_i \approx \widehat{\Sigma}_i$,
$\alpha_i(b_i) \approx \widehat{\alpha}_i$, and
$\varphi_i(\alpha_i(b_i)) \approx \widehat{\varphi}_i$
(on both segments and their concatenation), then $P_i$ transfers
by the same construction that guarantees well-definedness
in iterative specification.

Separately, validation criteria
$C_i = (C_\Sigma, C_\alpha, C_\varphi)$ specify the
comparison methods and tolerance thresholds used to evaluate
compatibility between $EC_i$ and $\widehat{EC}_i$.
These criteria are specified by the engineer and are not part
of the execution conditions themselves.

\vspace{1ex}
\paragraph{Verification.}

Verification evaluates formal correctness under specified execution
conditions:
\[
M_i, EC_i \models P_i.
\]

It depends on both the model $M_i$ and the execution conditions $EC_i$,
and checks semantic consistency, constraint satisfaction, and property
preservation.

\vspace{1ex}
\paragraph{Validation.}

Validation evaluates compatibility between specified and experimental
execution conditions:
\[
EC_i \;\text{vs}\; \widehat{EC}_i.
\]

Concretely, validation compares three components:
\[
\Sigma_i \;\text{vs}\; \widehat{\Sigma}_i,
\qquad
\alpha_i(b_i) \;\text{vs}\; \widehat{\alpha}_i,
\qquad
\varphi_i(\alpha_i(b_i)) \;\text{vs}\; \widehat{\varphi}_i,
\]
according to validation criteria $C_i$.
The constraint comparison is performed at both levels:
on individual activity segments (do local bounds hold?)
and on their concatenation (does cumulative admissibility hold?).

Properties $P_i$ are not directly compared. They are verified
formally on the model under $EC_i$. Validation ensures that the
execution conditions under which $P_i$ was proved are preserved
experimentally. If constraints hold on segments and on their
concatenation, and execution semantics are preserved, then $P_i$
transfers to execution. This is the bridge between verification
and validation in MSBE.

Validation is mediated by activity, which provides a common
level for aligning formal and experimental behaviors.

\vspace{1ex}
\paragraph{Execution Assessment.}

Execution is sustained when:
\begin{itemize}
\item[(i)] verification holds under $(M_i, EC_i)$,
\item[(ii)] validation holds between $EC_i$ and $\widehat{EC}_i$ according to $C_i$,
\item[(iii)] compatibility persists over a non-trivial time horizon.
\end{itemize}

To formalize the restriction of the admissible model space,
we define the following subsets of $\mathcal{M}_i$:
\[
\mathrm{Verif}(EC_i)
=
\{ M \in \mathcal{M}_i \mid M, EC_i \models P_i \},
\]
the set of models satisfying specified properties under
execution conditions $EC_i$, and
\[
\mathrm{Valid}(EC_i, \widehat{EC}_i, C_i)
=
\{ M \in \mathcal{M}_i \mid
\text{execution of } M \text{ yields }
\widehat{EC}_i \text{ compatible with } EC_i
\text{ according to } C_i \},
\]
the set of models whose experimental execution conditions
remain compatible with the specified ones, on both individual
activity segments and their concatenation.

The admissible model space is then restricted by intersection:
\[
\mathcal{M}_{i+1}
=
\mathcal{M}_i
\cap
\mathrm{Verif}(EC_i)
\cap
\mathrm{Valid}(EC_i, \widehat{EC}_i, C_i).
\]

All three experimental quantities
$\widehat{\Sigma}_i$, $\widehat{\alpha}_i$, and $\widehat{\varphi}_i$
are derived by evaluating the specified structures
on observed execution data. They do not constitute an independent
specification but the experimental mirror of the execution conditions
in $EC_i$, compared according to the validation criteria $C_i$.
When validation holds, the properties $P_i$ verified on the model
are warranted to hold in execution.

\vspace{1ex}
\paragraph{Iterative Specification of Execution Conditions and Model.}

If execution is not sustained, execution conditions are revised:

\[
EC_{i+1}
=
\mathrm{Specify}(EC_i, \widehat{EC}_i, C_i).
\]

This may affect:
\begin{itemize}
\item execution semantics $\Sigma_i$,
\item activity measure $\alpha_i$,
\item constraint functions $\varphi_i$,
\item specified properties $P_i$,
\item validation criteria $C_i$.
\end{itemize}

A new model is then specified under $EC_{i+1}$.

\vspace{1ex}
\paragraph{Executability (Stabilization).}

Executability corresponds to the fixed-point condition:

\[
EC_{i+1} = EC_i
\quad \text{and} \quad
\mathcal{M}_{i+1} = \mathcal{M}_i.
\]

At stabilization, execution conditions no longer require revision,
models remain valid under these conditions, and execution remains
coherent over time.

Executability therefore characterizes stabilization of execution
conditions and of the admissible model space they induce.

\vspace{1ex}
\paragraph{Summary of the Iterative Cycle.}

The execution-centered MSBE process follows a structured logical order.
First, execution conditions $EC_i = (\Sigma_i, \alpha_i, \varphi_i, P_i)$
and validation criteria $C_i$ are specified from requirements,
defining the semantic and constraint framework of execution.
Second, a model $M_i$ is specified under these conditions.
Third, formal and experimental executions produce behavior traces
from which experimental execution conditions
$\widehat{EC}_i = (\widehat{\Sigma}_i, \widehat{\alpha}_i,
\widehat{\varphi}_i)$ are derived.

Verification evaluates formal correctness of $(M_i, EC_i)$ under the
specified execution semantics. Validation compares $EC_i$ and
$\widehat{EC}_i$ according to $C_i$, assessing constraint satisfaction
on individual activity segments and on their concatenation.
Properties $P_i$ are not evaluated experimentally: when execution
conditions are preserved, verified properties transfer by construction.
Execution assessment determines whether coherent execution is sustained
over time. If not, execution conditions, validation criteria, and the
model are revised. Executability is reached when execution
conditions and the induced admissible model space stabilize under
sustained execution.

\vspace{1ex}
 
Table~\ref{tab:msbe-new-concepts} summarizes the new concepts introduced by MSBE and formalized in the cycle above.
 
\begin{table}[h!]
\caption{New MSBE concepts for CPS engineering.}
\label{tab:msbe-new-concepts}
\centering
\renewcommand{\arraystretch}{1.25}
\begin{tabularx}{\textwidth}{@{}X X@{}}
\toprule
\textbf{New concept} & \textbf{Engineering significance} \\
\midrule
 
\textbf{Executability}
& Temporal property ensuring sustained \textbf{coherent activity} under physical constraints \\
 
\textbf{Activity}
& Behavioral measure capturing \textbf{meaningful changes} and linking model semantics to execution \\
 
\textbf{Execution semantics as engineering task}
& Explicit specification of ordering, timing, and interaction rules governing execution on a given substrate \\
 
\textbf{Physical constraints as semantic boundary conditions}
& Integration of latency, energy, bandwidth, and causality limits into behavioral admissibility \\
 
\textbf{Iterative behavioral refinement mechanism}
& Systematic reduction of the admissible behavior space through successive constraint refinement \\
 
\textbf{Integrated Verification \& Validation}
& Alignment between \textbf{model-level verification} and \textbf{execution-level validation} \\
 
\bottomrule
\end{tabularx}
\end{table}

\section{Applying the Iterative Cycle to Representative Cyber-Physical System Classes}
\label{sec:classes}

This section instantiates the MSBE cycle on four classes of
cyber-physical systems. The classes are chosen to stress-test
different aspects of the framework.

Human-centric CPS (Section~\ref{subsec:human-cps}, human--robot
collaborative assembly) introduce stochastic execution
semantics: human behavior is unpredictable, and the activity
measure must capture interaction events whose timing is not
controlled. This class tests the framework under semantic
uncertainty.

Biophysical CPS (Section~\ref{subsec:biophysical-cps}, wildfire
propagation monitoring) introduce hybrid traces: continuous
physical dynamics (heat propagation) coexist with discrete events
(ignition thresholds). This class tests the framework on hybrid
activity measures and on the causal consistency of concatenated
segments.

Technological CPS (Section~\ref{subsec:technological-cps},
industrial robot controller) introduce hard real-time constraints:
the admissibility of each execution cycle is bounded by worst-case
execution times and bus latencies. This class tests the framework
under strict constraint evaluation where a single segment violation
invalidates the trace.

Digital twins (Section~\ref{subsec:digital-twin}, spacecraft
mission monitoring) introduce open-ended execution: the system
must sustain executability not for a bounded test but for the
entire operational lifetime. This class tests the framework on
long-term concatenation, cumulative drift, and the revision of
validation criteria $C_i$.

Together, the four classes cover the main dimensions along which
execution conditions vary: deterministic vs.\ stochastic 
semantics, continuous vs.\ discrete vs.\ hybrid traces, bounded
vs.\ open-ended execution horizons, and strict vs.\ tolerant
constraint evaluation. For each class, we specify the execution
conditions $EC_i = (\Sigma_i, \alpha_i, \varphi_i, P_i)$, describe
formal and experimental execution, illustrate verification and
validation, identify what triggers iterative refinement, and
characterize what executability means concretely.

\subsection{Human-Centric Cyber-Physical Systems}
\label{subsec:human-cps}

Consider a human--robot collaborative assembly cell where a robotic arm and a human operator jointly assemble components. The system must coordinate robot motion with human actions in real time.

\paragraph{Execution conditions.}
Execution semantics $\Sigma_i$ define an event-driven interaction protocol: the robot waits for detected human gestures before initiating complementary motions. State transitions are triggered by perception events (gesture recognition, proximity detection) rather than by uniform time steps.

The activity measure $\alpha_i$ (event count; see Appendix~\ref{app:activity}) extracts interaction-relevant segments from the joint behavior trace: gesture initiation by the human, robot motion onset, handover completion, and task-phase transitions. Raw sensor streams and low-level motor commands are filtered out.

Admissibility constraints $\varphi_i$ evaluate timing bounds on human--robot coordination. For instance, the delay between a detected handover gesture and robot grasp initiation must remain below a safety threshold. Constraint functions also evaluate spatial safety margins and attention-consistency conditions.

Specified properties $P_i$ include collision avoidance, task completion within bounded time, and preservation of interaction sequencing.

\paragraph{Formal and experimental execution.}
Formal execution $b_i = \mathrm{Exec}_f(M_i, \Sigma_i)$ simulates the interaction protocol using a model of human behavior (e.g., scripted gesture sequences with stochastic timing). The simulation produces expected activity traces: predicted handover timings, coordination delays, and task durations.

Experimental execution $\widehat{b}_i = \mathrm{Exec}_e(M_i)$ deploys the system with an actual human operator. Observed activity $\widehat{\alpha}_i$ is extracted from sensor logs as interaction segments (individual handovers, phase transitions). Experimental constraints $\widehat{\varphi}_i$ are measured on each segment (coordination delay of this handover) and on their concatenation (cumulative sequencing consistency across the full assembly task). Induced execution semantics $\widehat{\Sigma}_i$ reveal the effective event ordering, which may differ from the specified protocol due to human variability.

\paragraph{Verification and validation.}
Verification checks that the model $M_i$ under $EC_i$ satisfies $P_i$: does the specified interaction protocol guarantee collision avoidance and bounded task time for all admissible gesture sequences?

Validation compares specified and experimental execution conditions according to $C_i$. Are the observed coordination delays $\widehat{\varphi}_i$ consistent with the specified bounds $\varphi_i$ on each handover segment? Does the cumulative constraint evaluation over the full task trace remain admissible? Does the experimentally induced event ordering $\widehat{\Sigma}_i$ match the specified protocol $\Sigma_i$? If these conditions hold, the safety properties $P_i$ (collision avoidance, bounded task time) verified on the model transfer to execution. Discrepancies indicate that the human behavior model, the interaction protocol, or the constraint bounds require revision.

\paragraph{Iterative refinement and executability.}
If validation reveals that human response times systematically exceed the assumed bounds, the admissibility constraints $\varphi_i$ must be relaxed or the execution semantics $\Sigma_i$ must be revised to include adaptive waiting. This yields $EC_{i+1}$ and a new model $M_{i+1}$.

Executability is reached when the interaction protocol, the human behavior model, and the admissibility constraints stabilize: $EC_{i+1} = EC_i$ and $\mathcal{M}_{i+1} = \mathcal{M}_i$. The system can then sustain coherent collaborative execution across operators and sessions.

\subsection{Biophysical Cyber-Physical Systems}
\label{subsec:biophysical-cps}

Consider an environmental monitoring system tracking wildfire propagation. Distributed sensors measure temperature, humidity, and wind. A computational model predicts fire spread to support evacuation decisions.

\paragraph{Execution conditions.}
Execution semantics $\Sigma_i$ define a hybrid temporal regime. Continuous dynamics govern physical propagation (heat diffusion, wind-driven spread). Discrete transitions correspond to threshold crossings (ignition of a new cell, sensor alert triggering).

The activity measure $\alpha_i$ (hybrid: continuous propagation rate and discrete ignition events; see Appendix~\ref{app:activity}) extracts propagation-relevant events from the behavior trace: ignition events, front advancement between spatial cells, and regime transitions (e.g., from surface fire to crown fire). Continuous temperature variations within a cell below the ignition threshold are not activity.

Admissibility constraints $\varphi_i$ evaluate temporal and spatial consistency. Predicted front arrival times must fall within physically plausible bounds given wind speed and fuel load. Constraint functions also evaluate sensor update rates: if sensor data arrives too slowly relative to propagation speed, constraint satisfaction degrades.

Specified properties $P_i$ include causal consistency of propagation ordering (no cell ignites before its upwind neighbor), bounded prediction error relative to observations, and timely alert generation.

\paragraph{Formal and experimental execution.}
Formal execution simulates fire spread on a spatial grid under specified wind and fuel conditions. The simulation produces a sequence of predicted ignition events with associated times and locations.

Experimental execution receives real-time sensor data from the deployed network. Observed activity $\widehat{\alpha}_i$ consists of detected threshold crossings as individual segments. Experimental constraints $\widehat{\varphi}_i$ are evaluated on each propagation segment (delay between successive cell ignitions) and on their concatenation (global propagation ordering across the entire fire front). Induced semantics $\widehat{\Sigma}_i$ reflect the effective temporal and causal ordering of observed ignition events.

\paragraph{Verification and validation.}
Verification checks that the fire spread model under $EC_i$ preserves causal ordering and produces bounded prediction errors for the specified wind and fuel scenarios.

Validation compares predicted and observed propagation according to $C_i$. If the observed propagation order $\widehat{\Sigma}_i$ contradicts the specified causal ordering $\Sigma_i$, or if observed arrival times $\widehat{\varphi}_i$ systematically deviate from predicted bounds $\varphi_i$ on individual segments or on their concatenation, the execution conditions require revision. When segment and concatenation constraints are preserved, the properties $P_i$ (causal consistency, bounded prediction error) transfer from the verified model.

\paragraph{Iterative refinement and executability.}
Discrepancies may indicate that the spatial resolution of the model is too coarse, that the wind model is inadequate, or that sensor update rates are insufficient to track fast-moving fronts. Refinement updates the execution semantics $\Sigma_{i+1}$ (finer spatial discretization), the constraints $\varphi_{i+1}$ (revised timing bounds), or the activity measure $\alpha_{i+1}$ (new regime transition types).

Executability is reached when the model, the propagation semantics, and the sensor-driven constraint evaluation stabilize. The system can then sustain reliable fire tracking and alert generation under the specified environmental conditions.

\subsection{Technological Cyber-Physical Systems}
\label{subsec:technological-cps}

Consider an industrial robot controller executing a sequence of pick-and-place operations on a production line. The controller runs on an embedded processor and actuates a mechanical arm through a real-time communication bus.

\paragraph{Execution conditions.}
Execution semantics $\Sigma_i$ define a cyclic execution regime: at each cycle, the controller reads sensor inputs, computes a control law, and issues actuation commands. The cycle period is fixed by the real-time scheduler. State transitions follow a deterministic ordering within each cycle.

The activity measure $\alpha_i$ (event frequency per cycle; see Appendix~\ref{app:activity}) extracts operationally relevant events: grasp initiation, placement completion, mode transitions (idle to active, normal to error recovery), and configuration changes (tool switch, speed adaptation).

Admissibility constraints $\varphi_i$ enforce real-time bounds: the worst-case execution time of the control computation must not exceed the cycle period. Communication latency on the bus must remain below a specified threshold. Energy consumption per cycle must stay within the power budget.

Specified properties $P_i$ include deterministic actuation sequencing, bounded end-to-end latency from sensor reading to actuation, and absence of deadline misses.

\paragraph{Formal and experimental execution.}
Formal execution simulates the controller on a model of the execution platform, including the scheduler, the communication bus, and the mechanical dynamics. It produces expected activity traces: predicted cycle timings, actuation sequences, and resource consumption.

Experimental execution deploys the controller on the physical platform. Observed activity $\widehat{\alpha}_i$ is extracted from execution logs as individual cycle segments. Experimental constraints $\widehat{\varphi}_i$ are evaluated on each cycle (execution time, bus latency, power consumption) and on their concatenation over the operational period (cumulative resource consumption, absence of progressive timing degradation). Induced semantics $\widehat{\Sigma}_i$ reveal the effective execution ordering, which may differ from the specified cyclic regime due to interrupt handling, cache effects, or bus contention.

\paragraph{Verification and validation.}
Verification checks that the controller model under $EC_i$ satisfies $P_i$: does the specified cyclic regime guarantee absence of deadline misses and deterministic actuation for all admissible input sequences?

Validation compares specified and experimental execution conditions according to $C_i$. Do measured execution times $\widehat{\varphi}_i$ remain within the specified bounds $\varphi_i$ on each cycle? Does the concatenation of cycles show progressive degradation that individual cycles do not reveal? Does the effective execution ordering $\widehat{\Sigma}_i$ match the specified deterministic cycle $\Sigma_i$? If segment and concatenation constraints hold and semantics are preserved, the properties $P_i$ (deadline absence, deterministic sequencing) transfer. If bus contention introduces non-deterministic reordering, the specified semantics are not preserved and execution conditions require revision.

\paragraph{Iterative refinement and executability.}
If validation reveals systematic deadline overruns, the cycle period in $\Sigma_i$ may need to be increased, the control computation may need to be simplified (reducing $\mathcal{M}_i$), or the constraint bounds $\varphi_i$ may need to be tightened to exclude pathological bus contention scenarios. Each revision yields $EC_{i+1}$.

Executability is reached when the cyclic execution regime, the timing constraints, and the controller model stabilize. The system can then sustain deterministic real-time operation under the specified platform constraints.

\subsection{Digital Twins}
\label{subsec:digital-twin}

Consider a digital twin of a spacecraft used for mission monitoring. The digital twin maintains a synchronized computational replica of the spacecraft's thermal, power, and orbital subsystems. Ground operators use it to predict system evolution and plan maneuvers.

\paragraph{Execution conditions.}
Execution semantics $\Sigma_i$ define synchronized co-evolution between the physical spacecraft and its computational twin. The twin receives telemetry at bounded intervals and updates its internal state accordingly. Between updates, it extrapolates state evolution using its internal models. State transitions are triggered by telemetry reception, internal prediction steps, and operator commands.

The activity measure $\alpha_i$ (synchronization error rate; see Appendix~\ref{app:activity}) extracts mission-relevant events: orbital correction maneuvers, thermal mode transitions (eclipse entry/exit), power budget shifts, and anomaly detections. Routine telemetry updates that do not alter the mission-relevant state are not activity.

Admissibility constraints $\varphi_i$ evaluate synchronization quality. The deviation between the twin's predicted state and the received telemetry must remain below specified thresholds for each subsystem. Telemetry latency must not exceed the extrapolation validity horizon. Energy budget consistency between predicted and measured values must hold within specified margins.

Specified properties $P_i$ include bounded synchronization error, causal consistency of maneuver sequencing, and preservation of energy budget invariants.

\paragraph{Formal and experimental execution.}
Formal execution simulates the twin in isolation using a model of the spacecraft dynamics and a synthetic telemetry stream. It produces expected activity traces: predicted maneuver timings, thermal transitions, and synchronization error traces.

Experimental execution runs the twin against actual telemetry from the spacecraft. Observed activity $\widehat{\alpha}_i$ consists of detected mission events as individual segments. Experimental constraints $\widehat{\varphi}_i$ are evaluated on each update segment (synchronization error at this telemetry reception, latency of this update) and on their concatenation over mission phases (cumulative drift, long-term energy budget consistency). Induced semantics $\widehat{\Sigma}_i$ reflect the effective temporal ordering of state updates, which depends on communication delays and ground station availability.

\paragraph{Verification and validation.}
Verification checks that the twin model under $EC_i$ satisfies $P_i$: does the synchronization protocol guarantee bounded error and causal consistency for all admissible telemetry patterns?

Validation compares specified and experimental execution conditions over extended mission phases according to $C_i$. Do observed synchronization errors $\widehat{\varphi}_i$ remain within the specified bounds $\varphi_i$ on each update segment? Does the concatenation of updates over weeks or months reveal cumulative drift that individual segments do not? Does the effective update ordering $\widehat{\Sigma}_i$ preserve the causal structure assumed in $\Sigma_i$? If segment and concatenation constraints hold, the properties $P_i$ (bounded error, causal consistency, energy invariants) transfer. If ground station outages introduce long extrapolation gaps, the concatenation constraints may fail even when individual segment constraints are satisfied, triggering revision.

\paragraph{Iterative refinement and executability.}
If synchronization errors grow beyond bounds during eclipse phases, the thermal model may need refinement (restricting $\mathcal{M}_i$), the extrapolation horizon in $\varphi_i$ may need tightening, or the update semantics $\Sigma_i$ may need to account for variable telemetry availability. Refinement may also affect the validation criteria $C_i$ themselves. For instance, if eclipse transitions systematically cause transient synchronization spikes that do not compromise mission safety, $C_\varphi$ may be revised to tolerate larger deviations during eclipse entry and exit while tightening tolerances during nominal orbital phases. This illustrates that the validation criteria are engineering choices subject to iterative revision, not fixed thresholds.

Executability is reached when the twin's synchronization protocol, the subsystem models, and the admissibility constraints stabilize over mission duration. The twin can then sustain reliable state alignment with the physical spacecraft under evolving mission conditions. Digital twins thus amplify the executability challenge: stabilization must hold not for a bounded test but for the entire operational lifetime of the system.

\subsection{Synthesis Across Cyber-Physical System Classes}
\label{subsec:cps-synthesis}

Across these four classes, the MSBE cycle operates identically in structure. Execution conditions $EC_i = (\Sigma_i, \alpha_i, \varphi_i, P_i)$ are specified from domain requirements. Models are defined under these conditions. Formal and experimental executions produce behavior traces from which experimental execution conditions $\widehat{EC}_i = (\widehat{\Sigma}_i, \widehat{\alpha}_i, \widehat{\varphi}_i)$ are derived. Validation compares $EC_i$ and $\widehat{EC}_i$ according to criteria $C_i$, evaluating constraints on segments and on their concatenation. When validation holds, the properties $P_i$ verified on the model transfer to execution. When execution is not sustained, execution conditions and models are revised. Executability corresponds to stabilization.

What varies across classes is the content of each component. In human-centric CPS, $\Sigma_i$ is event-driven and $\alpha_i$ captures human actions. In biophysical CPS, $\Sigma_i$ is hybrid and $\alpha_i$ captures regime transitions. In technological CPS, $\Sigma_i$ is cyclic and $\alpha_i$ captures operational events. In digital twins, $\Sigma_i$ is synchronization-driven and $\alpha_i$ captures mission-relevant state changes. The constraints $\varphi_i$ range from coordination delays, to propagation bounds, to real-time deadlines, to synchronization errors. The properties $P_i$ range from safety and sequencing, to causal consistency, to bounded latency, to long-term alignment.

Table~\ref{tab:cps-ec-summary} summarizes the instantiation of execution conditions across CPS classes.

\begin{table}[h!]
\caption{Instantiation of execution conditions $EC_i = (\Sigma_i, \alpha_i, \varphi_i, P_i)$ across CPS classes.}
\label{tab:cps-ec-summary}
\centering
\renewcommand{\arraystretch}{1.25}
\begin{tabularx}{\textwidth}{@{}l X X X X@{}}
\toprule
\textbf{CPS class} & $\boldsymbol{\Sigma_i}$ & $\boldsymbol{\alpha_i}$ & $\boldsymbol{\varphi_i}$ & $\boldsymbol{P_i}$ \\
\midrule
Human-centric & Event-driven interaction protocol & Human actions, handovers, phase transitions & Coordination delays, safety margins & Collision avoidance, task completion \\
Biophysical & Hybrid continuous/ discrete regime & Ignition events, regime transitions & Propagation timing bounds, sensor update rates & Causal consistency, bounded prediction error \\
Technological & Cyclic real-time regime & Sensor updates, actuation commands, mode switches & Worst-case execution time, bus latency, energy & Deadline absence, deterministic sequencing \\
Digital twin & Synchronized co-evolution & Maneuvers, thermal transitions, anomalies & Synchronization error bounds, telemetry latency & Bounded error, causal consistency, energy invariants \\
\bottomrule
\end{tabularx}
\end{table}

The invariance of the cycle structure across these diverse instantiations confirms that the MSBE framework is not specific to a particular CPS domain. The four components of execution conditions, the separation between formal and experimental execution, the activity-mediated validation, and the stabilization criterion apply whenever system behavior depends on explicitly defined execution conditions. The framework is therefore applicable beyond CPS, to any engineered system where execution semantics must be treated as first-class engineering entities. CPS provide the motivating and most demanding context, but the formal apparatus generalizes to purely logical, purely experimental, or other classes of execution-constrained systems.

\section{Interdisciplinary Research Roadmap}
\label{sec:roadmap}

The MSBE perspective opens several research directions.
Table~\ref{tab:msbe-tms-roadmap} maps open CPS engineering gaps to
TMS foundations that can be developed within the MSBE framework.
 
\begin{table}[h!]
\caption{Open CPS engineering gaps and TMS foundations addressable within MSBE.}
\label{tab:msbe-tms-roadmap}
\centering
\renewcommand{\arraystretch}{1.25}
\begin{tabularx}{\textwidth}{@{}X X@{}}
\toprule
\textbf{Open gap in CPS engineering} & \textbf{TMS foundation to develop} \\
\midrule
 
Testing and validation remain scenario-dependent and tool-driven
& Validation criteria $C_i$ formalized as experimental frames \cite{traore2006ef,balci1998vv}, specifying objectives, assumptions, and constraints governing experimentation \\
 
Execution semantics are implicit or fixed at specification time
& Constraint-aware execution semantics integrating physical
bounds directly into $\Sigma_i$
\cite{zeigler2018tms,lee2008cps,lohstroh2021linguafranca} \\
 
Heterogeneous components yield emergent behavioral ambiguities
& Behavioral coherency under couplings: closure under composition \cite{zeigler2018tms,muzy2017iterative} \\
 
Simulation engines impose implicit ordering and timing assumptions
& Simulation as controlled execution with explicit semantic assumptions \cite{zeigler1976tms,zeigler2018tms} \\
 
System architectures evolve during operation
& Dynamic system structure integrated into behavioral modeling \cite{barros1997modeling,muzy2014specification,zeigler2018tms} \\
 
Verified properties do not transfer to physical execution
& Activity-mediated bridge between verification and validation (this paper) \\
 
\bottomrule
\end{tabularx}
\end{table}

\paragraph{Formalization of executability.}
Executability must be characterized as a temporal property over admissible behavior spaces under physical constraints.
This raises questions about invariants, stability criteria, and convergence guarantees of iterative behavioral refinement. The result that Finite and Timed Parallel DEVS can simulate any iterative specification \cite{zeigler2018tms} provides a computational bridge toward DEVS-based model checking. Connecting this result with multi-valued model-checking techniques \cite{chechik2003modelchecking} could enable automated verification of executability conditions.

\paragraph{Constraint-aware execution semantics.}
A systematic theory is required to integrate physical constraints (latency, bounded resources, energy, causality) directly into execution semantics rather than treating them as implementation details. Lee's call for rebuilding computing abstractions to embrace physical dynamics \cite{lee2008cps} remains largely unanswered at the formal level. Coordination languages such as Lingua Franca \cite{lohstroh2021linguafranca} make timing and determinism explicit at the programming level. Extending such approaches to account for substrate-level physical constraints within the MSBE framework is a concrete next step.

\paragraph{Bridging verification and validation.}
Future work must formalize the relation between model-level verification and execution-level validation, possibly through trace abstraction, activity-based metrics, and behavioral distance measures. The experimental frame methodology \cite{traore2006ef} provides a formal specification of the context governing experimentation and can serve as the foundation for specifying validation criteria $C_i$. Behavioral conformance checking, as developed for distributed service compositions \cite{poizat2016verchor}, offers a structural analogy: verifying that experimental execution conditions conform to specified ones is analogous to verifying that peer implementations conform to a choreography.

\paragraph{Hybrid behavioral reasoning.}
Cyber–physical systems combine discrete and continuous dynamics.
Developing scalable abstractions that preserve executability while enabling formal reasoning remains an open challenge. Deductive verification of hybrid programs \cite{platzer2018logical}, model checking of hybrid automata \cite{henzinger1996hybrid,alur2015cps}, and synchronous modeling of biophysical systems \cite{berry2000esterel,demaria2016lustre} each address parts of this challenge. Integrating these approaches within the MSBE cycle, where hybrid behavioral reasoning is grounded in explicit execution conditions and mediated by activity, is a key research direction.

\paragraph{Tool support for iterative system specification.}
Engineering environments must support explicit execution semantics, constraint refinement, and activity-based analysis in an integrated manner. Current industrial practice, as documented by Liebel et al.\ \cite{liebel2018sosym} and Baduel et al.\ \cite{baduel2018sysml}, shows persistent gaps between modeling environments and execution-aware verification. Bridging these gaps requires toolchains that integrate specification of execution conditions, formal and experimental execution, and iterative refinement within a single workflow.

\paragraph{Integration with language-level execution frameworks.}
Frameworks such as GEMOC \cite{combemale2014globalizing} and standards such as fUML/xMOF \cite{mayerhofer2013xmof} define and compose execution semantics at the modeling language level. At the programming level, coordination languages such as Lingua Franca \cite{lohstroh2021linguafranca} make timing, determinism, and concurrency semantics explicit for distributed CPS. A natural research direction is to apply the MSBE cycle to models and programs produced within these frameworks: given a system whose language-level or coordination-level execution semantics are explicitly defined, does the behavior produced by execution remain admissible when realized on a physically constrained CPS substrate? This would connect language-level executability (can this model be executed?), coordination-level determinism (are timing guarantees preserved?), and substrate-level executability (can execution be sustained under physical constraints?) within a single engineering workflow, and provide a formal basis for assessing when semantic guarantees are preserved or invalidated by deployment conditions.

\paragraph{Generalization beyond CPS.}
As discussed in Section~\ref{subsec:cps-synthesis}, the MSBE cycle generalizes to any system whose behavior depends on explicit execution conditions. Investigating this generalization requires identifying classes of non-CPS systems (purely logical, purely experimental, cognitive) where the framework applies, characterizing the role of physical constraints in each case, and determining whether activity-preserving morphisms can formalize behavioral equivalence across execution substrates.

\section{Conclusion}
\label{sec:conclusion}

This paper introduced Modeling and Simulation Based Engineering (MSBE) as a methodology for engineering cyber-physical systems under explicit execution semantics. The central thesis is that CPS behavior emerges from execution on physically constrained substrates, and that execution conditions must therefore be treated as first-class engineering entities rather than as implicit assumptions delegated to tools or platforms.

MSBE formalizes execution conditions as a tuple $EC_i = (\Sigma_i, \alpha_i, \varphi_i, P_i)$ and organizes engineering around an iterative cycle of formal execution, experimental execution, verification, validation, and refinement. Activity serves as the mediator between the formal and experimental domains, providing a common level at which behavioral admissibility is evaluated. Executability is defined as the stabilization of execution conditions and the induced admissible model space under sustained execution.

The framework was instantiated on four representative CPS classes. In each case, the same cycle structure applies while the content of execution conditions varies with the domain. This invariance supports the claim that the framework generalizes beyond CPS to any system whose behavior depends on explicitly defined execution conditions.

Several limitations should be acknowledged. The paper is a conceptual and methodological contribution. No tool implementation is provided, and the instanciations remain illustrative rather than experimentally validated. The formal apparatus defines what executability means but does not yet provide decidability results or complexity bounds for the stabilization criterion. The relation between the activity measure and existing notions of behavioral equivalence (bisimulation, trace equivalence) remains to be formalized.

These limitations point to concrete research directions. Formalizing executability as a temporal property with convergence guarantees would strengthen the mathematical foundations. Developing constraint-aware execution semantics that integrate physical constraints directly into formal models would enable automated verification under realistic assumptions. Defining activity-preserving abstractions would formalize cross-platform behavioral equivalence and support reuse of verification results across execution substrates. Tool support integrating explicit execution semantics, constraint refinement, and activity-based analysis is necessary for practical adoption.

More broadly, the work opens avenues for investigating behavioral preservation under system evolution, cross-substrate behavioral comparability, and the role of activity as a unifying measure across TMS, Software Engineering, Systems Engineering, and CPS research communities.

\appendix
\section{Formal Definitions of Activity Measures}
\label{app:activity}

This appendix provides the formal definitions of activity measures
used throughout the paper, adapted from the Theory of Modeling and
Simulation \cite{zeigler2018tms} and from \cite{muzy2019exploiting}.
It clarifies how activity measures relate to admissibility constraints
(on segments) and to specified properties (on traces).

\paragraph{Behavior traces, segments, and concatenation.}

A behavior trace\footnote{In TMS \cite{zeigler2018tms}, behavior traces are denoted $\omega_t$. We use $b$ throughout for consistency with the MSBE notation.} is a pair
$b = (b^{\mathrm{in}}, b^{\mathrm{out}})$, where
$b^{\mathrm{in}} : [0, t[ \;\rightarrow X \cup \{\phi\}$ is the input trace,
$b^{\mathrm{out}} : [0, t[ \;\rightarrow Y \cup \{\phi\}$ is the output trace,
$X$ and $Y$ are input and output value sets respectively,
and $\phi$ denotes a null value (no event).
In CPS, input traces typically carry sensor readings, telemetry,
or environmental stimuli; output traces carry actuation commands,
predictions, or control signals.
In the MSBE notation of Section~\ref{sec:cps-msbe},
a trace produced by formal execution is denoted
$b_i \in \mathcal{B}_{\Sigma_i}$, where $\mathcal{B}_{\Sigma_i}$
is the space of traces realizable under execution semantics $\Sigma_i$.
For readability, the body of the paper writes $b_i$ without
decomposing the input-output pair. The definitions below apply
to $b^{\mathrm{in}}$, $b^{\mathrm{out}}$, or both, depending
on the activity measure considered.

A \emph{segmentation} of a trace decomposes it into consecutive segments:
$b_i = b_i^{(1)} \bullet b_i^{(2)} \bullet \cdots \bullet b_i^{(n)}$,
where $b_i^{(k)}$ is the $k$-th segment defined on $[t_{k-1}, t_k[$
and $\bullet$ denotes concatenation.
This segmentation is central to the MSBE formalism:
admissibility constraints $\varphi_i$ are evaluated at two levels
(on individual segments and on their concatenation),
while specified properties $P_i$ hold on entire traces
(i.e., on the concatenation of all segments).

\paragraph{Continuous activity measure.}

For a continuous trace $b$ (Figure~\ref{fig:hybrid-trace}, continuous portions), the activity measure quantifies the rate of change between $0$ and $t$ \cite{muzy2019exploiting}:
\[
\alpha_i^{\mathrm{cont}}(b) = \int_0^t \left| \frac{db(\tau)}{d\tau} \right| \cdot d\tau.
\]

This measure corresponds to the total arc length of the trace.
It can be approximated by the sum of successive distances between
adjacent local extrema (marked by $\ast$ in Figure~\ref{fig:hybrid-trace}).
In the MSBE formalism, $\alpha_i^{\mathrm{cont}}$ is an instantiation
of $\alpha_i : \mathcal{B}_{\Sigma_i} \rightarrow \mathcal{A}_i$
for continuous execution semantics $\Sigma_i$.

\paragraph{Hybrid activity measure.}

For a hybrid trace $b$ combining continuous segments and discrete
discontinuities (Figure~\ref{fig:hybrid-trace}), the activity measure
is the total variation \cite{muzy2019exploiting}:
\[
\alpha_i^{\mathrm{hyb}}(b) =
\sum_{\ast} |\text{height between adjacent local extrema}|
\;+\;
\sum_{\text{disc.}} |\textcolor{red}{\text{height of the discontinuity}}|.
\]

The first sum captures continuous changes (via local extrema,
marked by $\ast$). The second sum captures discrete events
(via discontinuities, marked by dashed lines in Figure~\ref{fig:hybrid-trace}).
At each discontinuity, the extrema immediately before and after
contribute to both sums. Continuous values may be quantized by a
quantum $D$, yielding discretization schemes that compute the time
of quantum crossings. This hybrid measure is the most general: it
reduces to $\alpha_i^{\mathrm{cont}}$ when there are no discontinuities.

\paragraph{Discrete-event activity measure.}

For a discrete-event trace $b^e$ (Figure~\ref{fig:discrete-trace}),
where the trace alternates between null values $\phi$ (no event) and
event values (state changes), the activity measure can be instantiated as:
\begin{itemize}
\item the total variation of the trace
(sum of absolute differences between successive event values),
\item the event count $|\{k \mid b^e(t_k) \neq \phi\}|$ over $[0, t[$, or
\item the event frequency (event count divided by $t$).
\end{itemize}

Different choices yield different granularities for evaluating
behavioral admissibility. Event count is used in the human-centric
CPS example (Section~\ref{subsec:human-cps}); event frequency
per cycle in the technological CPS example (Section~\ref{subsec:technological-cps}).

\begin{figure}[t]
\centering
\begin{tikzpicture}[scale=0.85]
\draw[->] (0,0) -- (10,0) node[right] {$t$};
\draw[->] (0,0) -- (0,4.5) node[above] {$Z$};
 
\foreach \y in {1.0, 2.0, 3.0, 4.0}
  \draw[dotted, gray] (0,\y) -- (10,\y);
\draw[<->, gray, thin] (10.3,1.0) -- (10.3,2.0);
\node[right, gray, font=\scriptsize] at (10.3,1.5) {$D$};
 
\draw[thick, blue!70!black]
  (0,1.0) .. controls (0.8,3.2) and (1.6,3.8) .. (2.2,3.5)
  .. controls (2.6,3.2) and (3.0,1.5) .. (3.4,1.8)
  .. controls (3.8,2.1) and (4.2,3.0) .. (4.5,2.8);
 
\draw[thick, red, dashed] (4.5,2.8) -- (4.5,1.2);
\draw[thick, blue!70!black]
  (4.5,1.2) .. controls (5.0,0.8) and (5.5,1.0) .. (6.0,1.8)
  .. controls (6.5,2.6) and (7.0,3.2) .. (7.3,3.0);
 
\draw[thick, red, dashed] (7.3,3.0) -- (7.3,1.5);
\draw[thick, blue!70!black]
  (7.3,1.5) .. controls (7.8,1.2) and (8.5,2.0) .. (9.2,2.5);
 
\node[green!50!black, font=\large] at (0.0,0.7+0.2) {$\ast$};
\node[green!50!black, font=\large] at (1.9,3.4+0.2) {$\ast$};
\node[green!50!black, font=\large] at (3.3,1.75) {$\ast$};
\node[green!50!black, font=\large] at (4.5,2.6+0.25) {$\ast$};
\node[green!50!black, font=\large] at (4.5,1+0.2) {$\ast$};
\node[green!50!black, font=\large] at (7.3,3.0) {$\ast$};
\node[green!50!black, font=\large] at (7.3,1.4+0.2) {$\ast$};
\node[green!50!black, font=\large] at (9.2,2.5) {$\ast$};
 
\foreach \x/\lab in {0/0, 4.5/t_1, 7.3/t_2, 9.2/t_3}
  \node[below, font=\scriptsize] at (\x,0) {$\lab$};
 
\draw[<->, thick, gray] (0,-0.5) -- (4.5,-0.5)
  node[midway, below, font=\scriptsize] {$b^{(1)}$};
\draw[<->, thick, gray] (4.5,-0.5) -- (7.3,-0.5)
  node[midway, below, font=\scriptsize] {$b^{(2)}$};
\draw[<->, thick, gray] (7.3,-0.5) -- (9.2,-0.5)
  node[midway, below, font=\scriptsize] {$b^{(3)}$};
\end{tikzpicture}
\caption{Hybrid behavior trace $b_i = b_i^{(1)} \bullet b_i^{(2)} \bullet b_i^{(3)}$,
segmented at discontinuities $t_1$ and $t_2$.
Continuous portions are shown in blue.
Local extrema are marked by green asterisks ($\ast$),
including on both sides of each discontinuity;
discontinuities (discrete events) are shown by red dashed lines.
Horizontal dotted lines indicate quantization levels separated by quantum $D$.
The hybrid activity measure $\alpha_i^{\mathrm{hyb}}$ sums the heights
between adjacent extrema and the heights of discontinuities.
Admissibility constraints $\varphi_i$ evaluate each segment $b_i^{(k)}$
individually (local bounds) and their concatenation (global admissibility).}
\label{fig:hybrid-trace}
\end{figure}
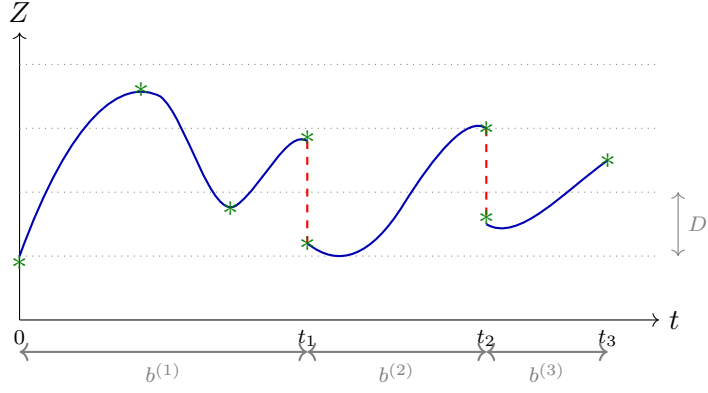

\begin{figure}[t]
\centering
\begin{tikzpicture}[scale=0.85]
\draw[->] (0,0) -- (10,0) node[right] {$t$};
\draw[->] (0,0) -- (0,3.5) node[above] {$Z \cup \{\phi\}$};
 
\node[left, font=\scriptsize, blue!70!black] at (0,0) {$\phi$};
 
\draw[thick, blue!70!black] (0,0) -- (1.0,0);
\draw[thick, orange!80!black] (1.0,0) -- (1.0,2.0);
\fill[orange!80!black] (1.0,2.0) circle (2pt);
\node[above, font=\scriptsize, orange!80!black] at (1.0,2.2) {change};
 
\draw[thick, blue!70!black] (1.0,0) -- (2.5,0);
\draw[thick, orange!80!black] (2.5,0) -- (2.5,1.3);
\fill[orange!80!black] (2.5,1.3) circle (2pt);
 
\draw[thick, blue!70!black] (2.5,0) -- (4.0,0);
\draw[thick, orange!80!black] (4.0,0) -- (4.0,2.5);
\fill[orange!80!black] (4.0,2.5) circle (2pt);
 
\draw[thick, blue!70!black] (4.0,0) -- (5.5,0);
 
\draw[thick, blue!70!black] (5.5,0) -- (6.5,0);
\draw[thick, orange!80!black] (6.5,0) -- (6.5,1.7);
\fill[orange!80!black] (6.5,1.7) circle (2pt);
 
\draw[thick, blue!70!black] (6.5,0) -- (8.0,0);
\draw[thick, orange!80!black] (8.0,0) -- (8.0,1.0);
\fill[orange!80!black] (8.0,1.0) circle (2pt);
 
\draw[thick, blue!70!black] (8.0,0) -- (9.5,0);
 
\node[below, font=\scriptsize] at (0,0) {$0$};
\node[below, font=\scriptsize] at (1.0,0) {$t_1$};
\node[below, font=\scriptsize] at (2.5,0) {$t_2$};
\node[below, font=\scriptsize] at (4.0,0) {$t_3$};
\node[below, font=\scriptsize] at (6.5,0) {$t_4$};
\node[below, font=\scriptsize] at (8.0,0) {$t_5$};
\node[below, font=\scriptsize] at (9.5,0) {$t_6$};
 
\draw[<->, thick, gray] (0,-0.5) -- (4.0,-0.5)
  node[midway, below, font=\scriptsize] {$b^{e(1)}$};
\draw[<->, thick, gray] (4.0,-0.5) -- (8.0,-0.5)
  node[midway, below, font=\scriptsize] {$b^{e(2)}$};
\draw[<->, thick, gray] (8.0,-0.5) -- (9.5,-0.5)
  node[midway, below, font=\scriptsize] {$b^{e(3)}$};
\end{tikzpicture}
\caption{Discrete-event behavior trace
$b_i^e = b_i^{e(1)} \bullet b_i^{e(2)} \bullet b_i^{e(3)}$,
segmented at $t_3$ and $t_5$.
The trace alternates between $\phi$ (no event, blue) and event values
(orange, at times $t_1, \ldots, t_5$).
The event count on segment $b_i^{e(1)}$ is 3; on $b_i^{e(2)}$
it is 2. Admissibility constraints $\varphi_i$ may bound the event
count or frequency per segment (local), while specified properties
$P_i$ hold on the full trace (e.g., total event ordering, bounded
cumulative latency across all segments).}
\label{fig:discrete-trace}
\end{figure}

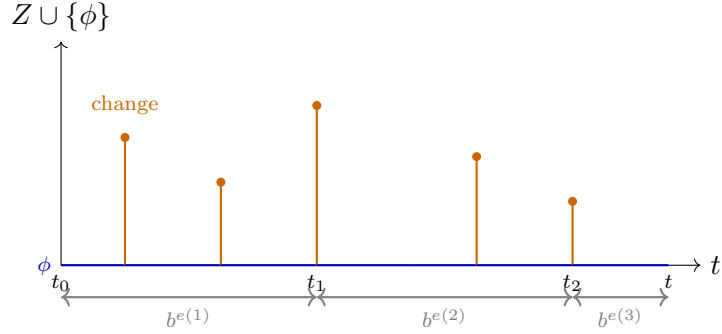
\begin{figure}[t]
\centering
\begin{tikzpicture}[scale=0.85]
\draw[->] (0,0) -- (10,0) node[right] {$t$};
\draw[->] (0,0) -- (0,3.5) node[above] {$Z \cup \{\phi\}$};

\node[left, font=\scriptsize, blue!70!black] at (0,0) {$\phi$};

\draw[thick, blue!70!black] (0,0) -- (1.0,0);
\draw[thick, orange!80!black] (1.0,0) -- (1.0,2.0);
\fill[orange!80!black] (1.0,2.0) circle (2pt);
\node[above, font=\scriptsize, orange!80!black] at (1.0,2.2) {change};

\draw[thick, blue!70!black] (1.0,0) -- (2.5,0);
\draw[thick, orange!80!black] (2.5,0) -- (2.5,1.3);
\fill[orange!80!black] (2.5,1.3) circle (2pt);

\draw[thick, blue!70!black] (2.5,0) -- (4.0,0);
\draw[thick, orange!80!black] (4.0,0) -- (4.0,2.5);
\fill[orange!80!black] (4.0,2.5) circle (2pt);

\draw[thick, blue!70!black] (4.0,0) -- (5.5,0);

\draw[thick, blue!70!black] (5.5,0) -- (6.5,0);
\draw[thick, orange!80!black] (6.5,0) -- (6.5,1.7);
\fill[orange!80!black] (6.5,1.7) circle (2pt);

\draw[thick, blue!70!black] (6.5,0) -- (8.0,0);
\draw[thick, orange!80!black] (8.0,0) -- (8.0,1.0);
\fill[orange!80!black] (8.0,1.0) circle (2pt);

\draw[thick, blue!70!black] (8.0,0) -- (9.5,0);

\foreach \x/\lab in {0/t_0, 4.0/t_1, 8.0/t_2, 9.5/t}
  \node[below, font=\scriptsize] at (\x,0) {$\lab$};

\draw[<->, thick, gray] (0,-0.5) -- (4.0,-0.5)
  node[midway, below, font=\scriptsize] {$b^{e(1)}$};
\draw[<->, thick, gray] (4.0,-0.5) -- (8.0,-0.5)
  node[midway, below, font=\scriptsize] {$b^{e(2)}$};
\draw[<->, thick, gray] (8.0,-0.5) -- (9.5,-0.5)
  node[midway, below, font=\scriptsize] {$b^{e(3)}$};
\end{tikzpicture}
\caption{Discrete-event behavior trace
$b_i^e = b_i^{e(1)} \bullet b_i^{e(2)} \bullet b_i^{e(3)}$,
segmented at $t_1$ and $t_2$.
The trace alternates between $\phi$ (no event, blue) and event values
(orange). The event count on segment $b_i^{e(1)}$ is 3; on $b_i^{e(2)}$
it is 2. Admissibility constraints $\varphi_i$ may bound the event
count or frequency per segment (local), while specified properties
$P_i$ hold on the full trace (e.g., total event ordering, bounded
cumulative latency across all segments).}
\label{fig:discrete-trace}
\end{figure}

\paragraph{Segments, traces, constraints, and properties.}

The distinction between segments and traces is central to the
MSBE formalism. Given a segmentation
$b_i = b_i^{(1)} \bullet b_i^{(2)} \bullet \cdots \bullet b_i^{(n)}$
of a behavior trace $b_i \in \mathcal{B}_{\Sigma_i}$:

\begin{itemize}
\item The activity measure $\alpha_i$ is evaluated on each segment
$b_i^{(k)}$ individually, producing segment-level activity values
$\alpha_i(b_i^{(k)}) \in \mathcal{A}_i$.

\item Admissibility constraints $\varphi_i$ operate at \emph{two levels}:
\begin{itemize}
\item \textit{On individual segments:}
$\varphi_i(\alpha_i(b_i^{(k)})) \in \mathcal{C}_i$ for each $k$.
This evaluates local bounds (e.g., timing of a single coordination
delay, resource usage within one execution cycle, synchronization
error at one telemetry update).
\item \textit{On their concatenation:}
$\varphi_i(\alpha_i(b_i^{(1)} \bullet \cdots \bullet b_i^{(n)})) \in \mathcal{C}_i$.
This evaluates global admissibility (e.g., cumulative drift,
preservation of causal ordering across successive segments,
absence of temporal blocking, progressive timing degradation).
\end{itemize}

\item Specified properties $P_i$ hold on \emph{entire traces}, i.e.,
on the concatenation
$b_i = b_i^{(1)} \bullet \cdots \bullet b_i^{(n)}$.
Properties are global by nature: safety, liveness, bounded error,
deterministic sequencing. They are verified on the model under
$EC_i$. They transfer to execution when $\varphi_i$ holds at
\emph{both} levels: if all segments satisfy local constraints and
their concatenation satisfies global admissibility, and execution
semantics $\Sigma_i$ are preserved, then $P_i$ transfers by
construction (Section~\ref{sec:cps-msbe}).
\end{itemize}

This two-level structure is anchored in the iterative specification
formalism \cite{muzy2017iterative,zeigler2018tms}: a behavior trace
is well-defined when all segments satisfy local constraints and their
concatenation satisfies global admissibility. The activity measure
determines the granularity of both evaluations.

\newpage
\bibliography{references}

\end{document}